\documentclass{article}
\usepackage{subfigure}
\usepackage{graphicx}
\usepackage{url,hyperref}
\usepackage{amsthm,amsmath}

\newtheorem{lemma}{Lemma}

\newtheorem{example}{Example}

\title{Interactive Constrained Association Rule Mining\footnote{A preliminary report on this work was presented
at the Second International Conference on Knowledge Discovery and
Data Mining~\cite{dawak00}.}}

\author{Bart Goethals\footnote{This work was done while the author was employed by the University of Limburg} \\
Helsinki Institute for Information Technology \\
\\
Jan Van den Bussche \\ 
University of Limburg}
\date{\ }

\begin{document}

\maketitle

\begin{abstract}
We investigate ways to support interactive mining sessions, in the
setting of association rule mining.  In such sessions, users
specify conditions (queries) on the associations to be generated.
Our approach is a combination of the integration of querying
conditions inside the mining phase, and the incremental querying
of already generated associations. We present several concrete
algorithms and compare their performance.
\end{abstract}

\section{Introduction}

The interactive nature of the mining process has been acknowledged
from the start~\cite{kddprocess}. It motivated the idea of a
``data mining query language''
\cite{han_dmql,dbminer,imiman_cacm,msql,ceri_sqlmine} and was
stressed again by Ng, Lakshmanan, Han and Pang~\cite{nlhp_car}. A
data mining query language allows the user to ask for specific
subsets of association rules by specifying several constraints
within each query.

In this paper, working in the concrete setting of association rule
mining, we consider a class of conditions on associations to be
generated which should be expressible in any reasonable data
mining query language: Boolean combinations of atomic conditions,
where an atomic condition can either specify that a certain item
occurs in the body of the rule or the head of the rule, or set a
threshold on the support or on the confidence. A \emph{mining
session} then consists of a sequence of such Boolean combinations
(henceforth referred to as \emph{queries}). Efficiently supporting
data mining query language environments is a challenging task.
Towards this goal, we present and compare three approaches. In the
first extreme, the \emph{integrated querying} approach, every
individual data mining query will be answered by running an
adaptation of the mining algorithm in which the constraints on the
rules and sets to be generated are directly incorporated. The
second extreme, the \emph{post-processing} approach, first mines
as much associations as possible, by performing one major, global
mining operation. After this relatively expensive operation, the
actual data mining queries issued by the user then amount to
standard lookups in the set of materialized associations. A third
approach, the \emph{incremental querying} approach, combines the
advantages of both previous approaches.

\subsection{Our contributions}

We present the first algorithm to support interactive mining
sessions efficiently.  We measure efficiency in terms of the total
number of itemsets that are generated, but do not satisfy the
query, and the number of scans over the database that have to be
performed. Specifically, our results are the following:

\begin{enumerate}

\item Although our results show significant improvements of
performance, we will also show that exploiting constraints is not
always the best solution.  More specifically, if mining without
constraints is feasible to begin with, then the presented
post-processing approach will eventually outperform integrated
querying.

\item The querying achieved by exploiting the constraints is
\emph{optimal}, in the sense that it never generates an itemset
that could give rise to a rule that does not satisfy the query,
apart from the minimal support and confidence thresholds.
Therefore, the number of generated itemsets during the execution
of a query, becomes proportional to the strength of the
constraints in the query: the more specific the query, the faster
its execution.

\item Not only is the number of passes trough the database
reduced, but also the size of the database itself, again
proportionally to the strength of the constraints in the query.

\item A generated itemset will, within a session, never be
regenerated as a candidate itemset:  results of earlier queries
are reused when answering a new query.

\end{enumerate}

This paper is further organized as follows. Section~\ref{related}
gives an overview of related work on constrained mining. In
Section~\ref{queries}, we present a way of incorporating
query-constraints inside a frequent set mining algorithm.  In
Section~\ref{interactive}, we discuss ways of supporting
interactive mining sessions. We conclude the paper in
Section~\ref{conclude}.

\section{Related Work}\label{related}

The idea that queries can be integrated in the mining algorithm
was initially launched by Srikant, Vu, and Agrawal~\cite{sva_car},
who considered queries that are Boolean expressions over the
presence or absence of certain items in the rules. Queries
specifically as bodies or heads were not discussed. The authors
considered three different approaches to the problem. The proposed
algorithms are not optimal: they generate and test several
itemsets that do not satisfy the query, and their optimizations
also do not always become more efficient for more specific
queries.

Also Lakshmanan, Ng, Han and Pang worked on the integration  of
constraints on itemsets in mining, considering conjunctions of
conditions on itemsets such as those considered here, as well as
others (arbitrary Boolean combinations were not
discussed)~\cite{lnhp_cfq,nlhp_car}.  Of the various strategies
for the so-called ``CAP'' algorithm they present, the one that can
handle the queries considered in the present paper is their
``strategy II''. Again, this strategy generates and tests itemsets
that do not satisfy the query.  Also, their algorithms implement a
rule-query by separately mining for possible heads and for
possible bodies, while we tightly couple the querying of rules
with the querying of sets. This work has also been further studied
by Pei, Han and Lakshmanan~\cite{pushmore,convertible}, and
employed within the FPgrowth algorithm.

Still other work focused on other kinds of constraints over
association rules and frequent sets, such as
\emph{correlation}~\cite{correlated}, and
\emph{improvement}~\cite{bag_constraint}. These and other
statistical measures of interestingness will not be discussed in
this paper.

All previously mentioned works do not discuss the reuse of results
acquired from earlier queries within a session.  Nag et al.
proposed the use of a knowledge cache for this
purpose~\cite{cache}.  Several caching strategies were studied for
different cache sizes.  However, their work only considers mining
sessions of queries where only constraints on the support of the
itemsets are allowed. No solutions were provided for other
constraints like those studied in this paper. Also Jeudy and
Boulicaut have studied the use of a knowledge cache for finding a
condensed representation of all itemsets, based on the concept of
\emph{free sets}~\cite{jeudy}.

\section{Review of the Apriori algorithm}

As introduced by Agrawal et al.~\cite{ais}, the association rule
mining problem can be described as follows: we are given a set of
items $\cal I$ and a database $\cal D$ of subsets of $\cal I$
called transactions. An association rule is an expression of the
form $B \Rightarrow H$, where $B$ and $H$ are sets of items
(itemsets). The \emph{support} of an itemset $I$ is the number of
transactions that include $I$. An itemset is called
\emph{frequent} if its support is no less than a given minimal
support threshold.  An association rule is called frequent if $B
\cup H$ is frequent and it is called \emph{confident} if the
support of $B \cup H$ divided by the support of $B$ exceeds a
given minimal confidence threshold.  The goal is now to find all
association rules over $\cal D$ that are frequent and confident.

The standard association rule mining algorithm
Apriori~\cite{kddboek_chap12} is divided in two phases: phase 1 generates
all frequent itemsets with respect to the given minimal support
threshold, and phase 2 generates all confident rules with respect
to the given minimal confidence threshold.

Phase 1 is performed based on the observation (also called the
anti-monotonicity property) that all supersets of an infrequent
itemset are also infrequent. An itemset is thus potentially
frequent, also called a \emph{candidate} itemset, if its support
is unknown and all of its subsets are frequent. In every step of
the algorithm, all candidate itemsets are generated and their
supports are then counted by performing a complete scan of the
transaction database. This is repeated until no new candidate
itemsets can be generated.

Phase 2 generates for every frequent itemset a set of rules by
dividing the itemset in potential bodies and heads. This can be
done in a similar level-wise manner as in phase 1, based on the
observation that if a head-set represents a confident rule for
that itemset, then all of its subsets also represent confident
rules~\cite{sa_general}. For example, if the itemset $\{1,2,3,4\}$
is a frequent set and $\{1,2\} \Rightarrow \{3,4\}$ is a confident
rule, then $\{1,2,3\} \Rightarrow \{4\}$ and $\{1,2,4\}
\Rightarrow \{3\}$ must also be confident. In every step within
phase 2, all candidate head-sets are generated and their
confidences are computed, until no new candidate head-sets can be
generated. Because we do not need to access the database, phase 2
is much faster in comparison with phase 1.

The performance of Apriori-like algorithms is highly dependent of
three factors:
\begin{enumerate}
\item the number of candidate patterns increases exponentially
with a decreasing minimal support threshold, \item the number of
association rules can become very large for small confidence
thresholds, and \item the size of the transaction database is
typically very large, such that scanning the database becomes a
costly operation.
\end{enumerate}
Also the length of the transactions (density of the database)
plays an important role, because large transactions can result in
large frequent patterns, implying a lot of candidate patterns and
a lot of scans through the database. Since the introduction of the
Apriori algorithm, a lot of research has been done to improve its
performance by improving on one or more of these factors. Almost
all improvements rely on its levelwise, bottom-up, breadth-first
nature and on the anti-monotonicity property of the minimal
support threshold.

Nevertheless, Han et al. presented the FPgrowth
algorithm~\cite{fpgrowth}, which uses a depth-first strategy.
Although this algorithm has a very efficient counting mechanism,
it suffers from two major deficiencies: \begin{enumerate} \item it
cannot exploit the anti-monotonicity property, resulting in a lot
more candidate patterns, and \item although the used trie data
structure somewhat compresses the transaction database, the
algorithm implicitly requires the database to reside in
main-memory.
\end{enumerate}

Although recent improvements have increased the performance of
Apriori tremendously, the support and confidence thresholds can
always be set low enough, resulting in an exponential blowup of
the number of patterns and rules. Nevertheless, such low
thresholds can still reveal interesting patterns and rules, but
one is not interested in all of the discovered patterns and rules,
but queries out the interesting ones according to some specified
constraints. Pushing these constraints as deep as possible into
the mining algorithm, such that the amount of computation is
proportional to what the user gets, should improve its performance
and allow lower thresholds.

\section{Exploiting Constraints}\label{queries}

As already mentioned in the Introduction, the constraints we
consider in this paper are Boolean combinations of atomic
conditions.  An atomic condition can either specify that a certain
item $i$ occurs in the body of the rule or the head of the rule,
denoted respectively by ${\rm Body}(i)$ or ${\rm Head}(i)$, or set
a threshold on the support or on the confidence.

In this section, we explain how we can incorporate these
constraints in the mining algorithm.  We first consider the
special case of constraints where only conjunctions of atomic
conditions or their negations are allowed.

\subsection{Conjunctive Constraints} \label{conjunct}

Let $b_1$, \ldots, $b_\ell$ be the items that must be in the body
by the constraint; $b'_1$, \ldots, $b'_{\ell'}$ those that must not;
$h_1$, \ldots, $h_m$ those that must be in the head; and $h'_1$,
\ldots, $h'_{m'}$ those that must not.

Recall that an association rule $X \Rightarrow Y$ is only
generated if $X \cup Y$ is a frequent set. Hence, we only have to
generate those frequent sets that contain every $b_i$ and $h_i$,
plus some of the subsets of these frequent sets that can serve as
bodies or heads.  Therefore we will create a set-query
corresponding to the rule-query, which is also a conjunctive
expression, but now over the presence or absence of an item $i$ in
a frequent set, denoted by ${\rm Set}(i)$ and $\neg {\rm Set}(i)$.
We do this as follows:
\begin{enumerate}

\item For each positive literal ${\rm Body}(i)$ or ${\rm Head}(i)$ in
the rule-query, add the literal ${\rm Set}(i)$ in the set-query.

\item If for an item $i$ both $\neg {\rm Body}(i)$
and $\neg {\rm Head}(i)$ are in the rule-query, add the negated
literal $\neg {\rm Set}(i)$ to the set-query.

\item Add the minimal support threshold to the set-query.

\item All other literals in the rule-query are ignored because they
do not restrict the frequent sets that must be generated.

Formally, the following is readily verified:

\end{enumerate}

\begin{lemma} \label{lemma1}
An itemset $Z$ satisfies the set-query if and only if there
exists itemsets $X$ and $Y$ such that $X \cup Y = Z$ and the rule
$X \Rightarrow Y$ satisfies the rule-query, apart from the
confidence threshold. \qed
\end{lemma}

So, once we have generated all sets $Z$ satisfying the set-query,
we can generate all rules satisfying the rule-query by splitting
all these $Z$ in all possible ways in a body $X$ and a head $Y$
such that the rule-query is satisfied. Lemma~\ref{lemma1}
guarantees that this method is ``sound and complete''.

So, we need to explain two things:
\begin{enumerate}
\item Finding all frequent $Z$ satisfying the set-query.
\item Finding, for each such $Z$, the frequencies of all bodies
and heads $X$ and $Y$ such that $X \cup Y = Z$ and $X \Rightarrow
Y$ satisfies the rule-query.
\end{enumerate}

\paragraph{Finding the frequent sets satisfying the set-query}
Let ${\it Pos} := \{i \mid {\rm Set}(i)$ in set-query$\}$ and
${\it Neg} := \{i \mid \neg{\rm Set}(i)$ in set-query$\}$. Note
that ${\it Pos} = \{b_1,\ldots,b_{\ell},h_1,\ldots,h_m\}$. Denote
the dataset of transactions by ${\cal D}$. We define the following
derived dataset ${\cal D}_0$: $${\cal D}_0 := \{t-({\it Pos} \cup
{\it Neg}) \mid \mbox{$t \in {\cal D}$ and ${\it Pos} \subseteq
t$}\}$$ In other words, we ignore all transactions that are not
supersets of ${\it Pos}$ and from all transactions that are not
ignored, we remove all items in ${\it Pos}$ plus all items that
are in ${\it Neg}$.

We observe:

\begin{lemma} \label{lemma2}
Let $p$ be the support threshold defined in the query. Let ${\cal
S}_0$ be the set of itemsets over the new dataset ${\cal D}_0$,
without any further conditions, except that their support is at
least $p$. Let $\cal S$ be the set of itemsets over the original
dataset $\cal D$ that satisfy the set-query, and whose support is
also at least $p$. Then $$ {\cal S} = \{s \cup {\it Pos} \mid s
\in {\cal S}_0\}. $$
\end{lemma}
\begin{proof}
To show the inclusion from left to right, consider $Z \in {\cal S}$.
We show that $s := Z - {\it Pos}$ is in ${\cal S}_0$.  Thereto, it suffices
to establish
an injection $t \mapsto t_0$
from the transactions $t$ in the support set of $Z$ in $\cal D$ (i.e., the set of all
transactions in ${\cal D}$ containing $Z$) into the transactions $t_0$ in
the support set of $s$ in ${\cal D}_0$.

Let $t$ be in ${\cal D}$ and containing $Z$.  Since $Z$ satisfies the
set-query, $Z$ contains ${\it Pos}$, and hence $t$ contains ${\it Pos}$
as well.
Thus, $t_0 := t - ({\it Pos} \cup {\it Neg})$ is in ${\cal D}_0$.
Since $Z \cap {\it Neg} = \emptyset$ (again because $Z$ satisfies the
set-query), $t_0$ contains $Z - {\it Pos} = s$.
Hence, $t_0$ is in the support set of $s$ in ${\cal D}_0$, as desired.

To show the inclusion from right to left, consider $s \in {\cal
S}_0$.  We show that $Z := s \cup {\it Pos}$ is in ${\cal S}$.
Thereto, it suffices to establish an injection $t_0 \mapsto t$
from the transactions $t_0$ in the support set of $s$ in ${\cal
D}_0$ into the transactions $t$ in the support set of $Z$ in $\cal
D$.

Let $t_0$ be in ${\cal D}_0$ and containing $s$. Obviously, a
transaction $t \in {\cal D}$ exists, such that $t_0 \cup {\it Pos}
\subseteq t - {\it Neg} \subseteq t$. Since $t$ contains $s \cup
{\it Pos}$, it is in the support set of $Z$ in ${\cal D}$, as
desired.
\end{proof}

We can thus perform any frequent set generation algorithm, using
only ${\cal D}_0$ instead of ${\cal D}$.  Note that the number of
transactions in ${\cal D}_0$ is exactly the support of ${\it Pos}$
in ${\cal D}$. Also, the search space of all itemsets is halved
for every item in ${\it Pos} \cup {\it Neg}$. In practice, the
search space of all frequent itemsets is at least halved for every
item in ${\it Pos}$ and at most halved for every item in ${\it
Neg}$. Still put differently: we are mining in a world where
itemsets that do not satisfy the query simply do not exist. The
correctness and optimality of our method is thus automatically
guaranteed.

Note however that now an itemset $I$, actually represents the
itemset $I \cup {\it Pos}$! We thus head-start with a lead of $k$,
where $k$ is the cardinality of ${\it Pos}$, in comparison with
standard, non-constrained mining.

\paragraph{Finding the frequencies of bodies and heads}
We now have all frequent sets containing every $b_i$ and $h_i$,
from which rules that satisfy the rule-query can be generated.
Recall that in the standard association rule mining algorithm
rules are generated by taking every item in a frequent set as a
head and the others as body.  All heads that result in a confident
rule, with respect to the minimal confidence threshold, can then
be combined to generate more general rules.  But, because we now
only want rules that satisfy the query, a head must always be a
superset of $\{h_1,\ldots,h_m\}$ and may not include any of the
$h'_i$ and $b_i$ (the latter because bodies and heads of rules are
disjoint). In this way, we head-start with a lead of $m$.
Similarly, a body must always be a superset of
$\{b_1,\ldots,b_{\ell}\}$ and may not include any of the $b'_i$
and $h_i$.

The following lemma (which follows immediately from Lemma~\ref{lemma2})
tells us that these potential heads and bodies
are already present, albeit implicitly, in ${\cal S}_0$:

\begin{lemma} \label{lemma3}
Let ${\cal S}_0$ be as in Lemma~\ref{lemma2}.  Let ${\cal B}$
(${\cal H}$) be the set of bodies (heads) of those association
rules over ${\cal D}$ that satisfy the rule-query. Then
$${\cal B} = \{s \cup \{b_1,\ldots,b_{\ell}\} \mid \mbox{$s \in {\cal S}_0$ and $s \cap \{b'_1,\ldots,b'_{\ell'},h_1,\ldots,h_m\} = \emptyset$} \} $$
and
$$ {\cal H} = \{s \cup \{h_1,\ldots,h_m\} \mid \mbox{$s \in {\cal S}_0$ and $s \cap \{h'_1,\ldots,h'_{m'}\,b_1,\ldots,b_{\ell}\} = \emptyset$} \}. $$
\end{lemma}

So, for the potential bodies (heads), we use, in ${\cal S}_0$, all
sets that do not include any of the $b'_i$ and $h_i$ ($h'_i$ and
$b_i$), and add all $b_i$ ($h_i$).  Hence, all we have to do is to
determine the frequencies of these subsets by performing one
additional scan trough the dataset. (We do not necessarily yet
have these frequencies because these sets do not contain either
items $b_i$ or $h_i$, while we ignored transactions that did not
contain all items $b_i$ and $h_i$.)

Each generated itemset can thus have up to three different
``personalities:''

\begin{enumerate}
\item A frequent set that satisfies the set-query;
\item A frequent set that can act as body of a rule that satisfies the
rule-query;
\item A frequent set that can act as head of a rule that satisfies the
rule-query.
\end{enumerate}

Hence, we finally have at most three families of sets, i.e., those
sets from which rules must be generated, the \emph{rule-sets}
(${\cal S}_0$ with all $b_i$ and $h_i$ added); a family of
possible bodies, the \emph{body-sets} (${\cal S}_0$ with all $b_i$
added, minus all those sets that include any of the $b'_i$ and
$h_i$); and yet another family of possible heads, the
\emph{head-sets} (${\cal S}_0$ with all $h_i$ added, minus all
those sets that include any of the $h'_i$ and $b_i$). Note that
the frequencies of the body-sets and head-sets need not necessarily to
be recounted since their frequencies are equal to the frequencies of their
corresponding sets in ${\cal S}_0$ if the query consists
of negated atoms only.
We finally generate the
desired association rules from the rule-sets, by looking for
possible bodies and heads only within the body-sets and head-sets
respectively, on condition that they have enough confidence.

\paragraph{Optimality}
Note that every rule-set, body-set, and head-set
is needed to construct the rules potentially satisfying
the rule-query so that these can be tested for confidence,
and moreover, no other sets are ever needed.
In this precise sense, our method is \emph{optimal}.

\begin{example}
We illustrate our method with an example. Assume we are given the
rule-query
\begin{multline*}
{\rm Body}(1) \wedge \neg {\rm Body}(2) \wedge {\rm
Head}(3) \wedge \neg {\rm Head}(4) \\
{} \wedge \neg {\rm Body}(5) \wedge \neg {\rm Head}(5) \wedge {\rm support} \geq 1 \wedge {\rm
confidence} \geq 50\%.
\end{multline*}
We begin by converting it to the
set-query $${\rm Set}(1) \wedge {\rm Set}(3) \wedge \neg {\rm
Set}(5) \wedge {\rm support} \geq 1.$$ Hence ${\it Pos} = \{1,3\}$
and ${\it Neg} = \{5\}$. Consider a database consisting of the
three transactions $\{2,3,5,6,9\}$, $\{1,2,3,5,6\}$ and
$\{1,3,4,8\}$. We ignore the first transaction because it is not a
superset of ${\it Pos}$.  We remove items $1$ and $3$ from the
second transaction because they are in ${\it Pos}$, and we also
remove $5$ because it is in ${\it Neg}$. We only remove items $1$
and $3$ from the third transaction. Table~\ref{vb1} shows the
itemsets that result from the mining algorithm after reading,
according to Lemma~\ref{lemma1} and~\ref{lemma2}, the two
resulting transactions. For example, the itemset $\{4,8\}$
actually represents the set $\{1,3,4,8\}$. It also represents a
potential body, namely $\{1,4,8\}$, but it does not represent a
head, because it includes item $4$, which must not be in the head
according to the given rule-query.  As another example, the empty
set now represents the set $\{1,3\}$ from which a rule can be
generated. It also represents a potential body and a potential
head.
\end{example}

\begin{table}
\centering
\begin{tabular}{|c||c||c|c|}
\hline
${\cal S}_0$ & ${\cal S}$ & ${\cal B}$ & ${\cal H}$ \\
\hline
\hline
$\{\}$ & $\{1,3\}$ & $\{1\}$ & $\{3\}$ \\
\hline
$\{2\}$ & $\{1,2,3\}$ & - & $\{2,3\}$ \\
\hline
$\{4\}$ & $\{1,3,4\}$ & $\{1,4\}$ & - \\
\hline
$\{6\}$ & $\{1,3,6\}$ & $\{1,6\}$ & $\{3,6\}$ \\
\hline
$\{8\}$ & $\{1,3,8\}$ & $\{1,8\}$ & $\{3,8\}$ \\
\hline
$\{2,6\}$ & $\{1,2,3,6\}$ & - & $\{2,3,6\}$ \\
\hline
$\{4,8\}$ & $\{1,3,4,8\}$ & $\{1,4,8\}$ & - \\
\hline
\end{tabular}
  \caption{An example of generated sets, which can represent a frequent set, as well as a body, as well as a head}
  \label{vb1}
\end{table}

\subsection{Boolean Queries}

Assume now given a rule-query that is an arbitrary Boolean
combination of atomic conditions.  We can put it in disjunctive
normal form and then generate all frequent itemsets for every disjunct
(which is a conjunction) in parallel by feeding every transaction
of the database to every disjunct, and processing them there as
described in the previous subsection.

However, this approach is a bit simplistic, as it might generate
some sets and rules multiple times. For example, consider the
following query: ${\rm Body}(1) \vee {\rm Body}(2)$. If we
convert it to its corresponding set-query (disjunct by disjunct),
we get ${\rm Set}(1) \vee {\rm Set}(2)$. Then, we would generate
for both disjuncts all supersets of $\{1,2\}$. We can avoid this
problem by putting the set-query to \emph{disjoint
DNF}.\footnote{In disjoint DNF, the conjunction of any two
disjuncts is unsatisfiable. Any boolean expression has an
equivalent disjoint DNF.} Then, no itemset can satisfy more than
one set-disjunct. On the other hand this does not solve the
problem of generating some rules multiple times. Consider the
equivalent disjoint DNF of the above set-query: ${\rm Set}(1)
\vee ({\rm Set}(2) \wedge \neg {\rm Set}(1))$. The first disjunct
thus contains the set $\{1,2\}$ and all of its supersets. If we
generate for every itemset all potential bodies and heads
according to every rule-disjunct, both rule-disjuncts will still
generate all rules with the itemset $\{1,2\}$ in the body. The
easiest way to avoid this problem is to put already the
rule-query in disjoint DNF. Obviously, this does not mean its
corresponding set-query is also in disjoint DNF, and hence, we
still have to put it in disjoint DNF.

After all sets have been generated according to the set-query, we
still have to generate all rules according to the rule-query.
This can be done for every rule-disjunct (which is a conjunction)
in parallel after some modifications to the algorithm described in
the previous subsection.

Indeed, a single set-disjunct can now contain sets from which
rules can be generated satisfying several rule-disjuncts. Hence,
a set generated in one set-disjunct has now possibly even more
personalities. More specifically, it can possibly represent for
every rule disjunct a set from which rules can be generated, a
body of a such a rule and a head of such a rule.  We illustrate
this with the rule-query given in the previous paragraph.
\begin{example} Assume we are given the rule-query $${\rm Body}(1)
\wedge ({\rm Body}(2) \vee {\rm Head}(2)).$$  In disjoint DNF,
this gives
$$({\rm Body}(1) \wedge {\rm Body}(2)) \vee ({\rm Body}(1) \wedge
{\rm Head}(2)).$$ Converted to its corresponding set-query in
disjoint DNF, we get $${\rm Set}(1) \wedge {\rm Set}(2).$$
Obviously, this single set-disjunct contains sets from which rules
satisfying the first rule-disjunct can be generated. Following the
methodology described in the previous subsection, this means we
still have to count the frequencies of all these sets without item
$1$ and item $2$ included, since they will occur as heads in the
rules satisfying the first rule disjunct. But now, the
set-disjunct also contains sets from which rules satisfying the
second rule-disjunct can be generated. Hence, we still have to
count the frequencies of the generated sets with item $1$
included, which can serve as bodies for the rules satisfying the
second rule-disjunct, and the sets with item $2$ included, which
can serve as heads.
\end{example}

Until now, we have disregarded the possible presence of negated
thresholds in the queries, which can come from the conversion to
disjoint DNF, or from the user himself.  In the latter case, it
would not be possible to exploit this constraint in an
Apriori-like algorithm, because it is an essentially bottom-up
algorithm. Algorithms that generate sets also in a top-down
strategy could exploit this constraint. Another source for negated
thresholds is the conversion from the user's query to a Disjoint
DNF formula. Before we discuss this, we first have to explain how
we are going to convert a given formula to disjoint DNF.

We first put the Boolean expression $\phi$ in DNF, obtaining an
expression of the form $\phi_1 \vee \phi_2 \vee \cdots \vee
\phi_n$, in which $\phi_i$ is a conjunction of atomic conditions
or their negations. Of course, any two of these disjuncts may not
be disjoint. A good way to obtain a disjoint DNF is to add to
every disjunct $\phi_i$ the negated disjuncts $\phi_j$ with $j <
i$. We thus become the equivalent formula $\phi_1 \vee (\phi_2
\wedge \neg \phi_1) \vee \cdots \vee (\phi_n \wedge \neg
\phi_{n-1} \wedge \cdots \wedge \neg \phi_1)$ in which all
disjuncts are pairwise disjoint. Our problem is not yet solved,
because our formula is not even in DNF anymore. We thus still will
have to convert every disjunct on itself to disjoint DNF. For
example, take $(\phi_2 \wedge \neg \phi_1)$ with $\phi_1 \equiv
p_1 \wedge p_2 \wedge \cdots \wedge p_\ell$ in which $p_i$ is an
atomic condition or its negation. The disjunct thus  becomes
$(\phi_2 \wedge \neg p_1) \vee (\phi_2 \wedge p_1 \wedge \neg p_2)
\vee \cdots \vee (\phi_2 \wedge p_1 \wedge p_2 \wedge \cdots
\wedge p_{\ell-1} \wedge \neg p_\ell)$, which is in disjoint DNF.

An example showing that negated thresholds can be introduced in
this process, is the following. \begin{example} Assume we are
given the rule-query $$({\rm Body}(1) \wedge {\rm support} \geq
10) \vee ({\rm Body}(2) \wedge {\rm support} \geq 5).$$ As
equivalent disjoint DNF, we obtain
\begin{multline*} ({\rm Body}(1) \wedge {\rm support} \geq 10)
{} \vee ({\rm Body}(2) \wedge {\rm support} \geq 5 \wedge \neg
{\rm Body}(1)) \\ {} \vee ({\rm Body}(2) \wedge {\rm support} \geq
5 \wedge {\rm Body}(1) \wedge {\rm support} < 10).
\end{multline*} Notice the maximal support threshold in the last
disjunct, which is needed to avoid generating itemsets satisfying
${\rm Body}(2) \wedge {\rm support} \geq 10 \wedge {\rm Body}(1)$
which are already generated by the first disjunct.
\end{example}

Negated support thresholds can be avoided however. After putting
the user's formula in DNF, but before putting the DNF in disjoint
DNF, we sort all disjuncts on their support threshold, in
ascending order. This guarantees that the conversion to disjoint
DNF does not introduce any negated support thresholds.

Note that we cannot avoid negated \emph{confidence} thresholds at
the same time: we have already sorted on support, and thus cannot
sort anymore on confidence at the same time. Since we are here
already in phase 2, it is less of an efficiency issue to just
ignore maximal confidence thresholds.

Furthermore, if a set-disjunct (rule-disjunct) consists of nothing
but a negated support (confidence) threshold, we can of course
easily switch the generation algorithm and generate the candidate
sets (heads) in a top-down manner.

\subsection{Experiments}

For our experiments, we have implemented an extensively optimized
version of the Apriori algorithm, equipped with the querying
optimizations as described in the previous sections.

We have experimented using three real data sets, of which two are
publicly available and one synthetic data set generated by the
program provided by the Quest research group at IBM
Almaden~\cite{datagenerator}. The mushroom data set contains
characteristics of various species of mushrooms, and was
originally obtained from the {UCI} repository of machine learning
databases~\cite{ucimlr}. The BMS-WebView-1 data set contains
several months worth of click-stream data from an e-commerce web
site, and is made publicly available by Blue Martini
Software~\cite{kddcup2000}. The basket data set contains
transactions from a Belgian retail store, but can unfortunately
not be made publicly available. Table~\ref{database} shows the
number of items and the number of transactions in each data set.
The table additionally shows the minimal support threshold we used
in our experiments for each data set, together with the resulting
number of iterations and the time (in seconds) which the Apriori
algorithm needed to find all frequent patterns.
\begin{table}
\centering
\begin{tabular}{|l|c|c|c|c|c|}
  \hline
  Data set & \#Items & \#Transactions & MinSup & It's & Time \\
  \hline
    T40I10D100K & 1\,000 & 100\,000 & 700 & 18 & 1\,700s \\
    mushroom & 120 & 8\,124 & 813 & 16 & 663s \\
    BMS-Webview-1 & 498 & 59\,602 & 36 & 15 & 86s \\
    basket & 13\,103 & 41\,373 & 5 & 11 & 43s \\
  \hline
\end{tabular}
\caption{Data set Characteristics} \label{database}
\end{table}

\begin{figure}
\centering
\includegraphics{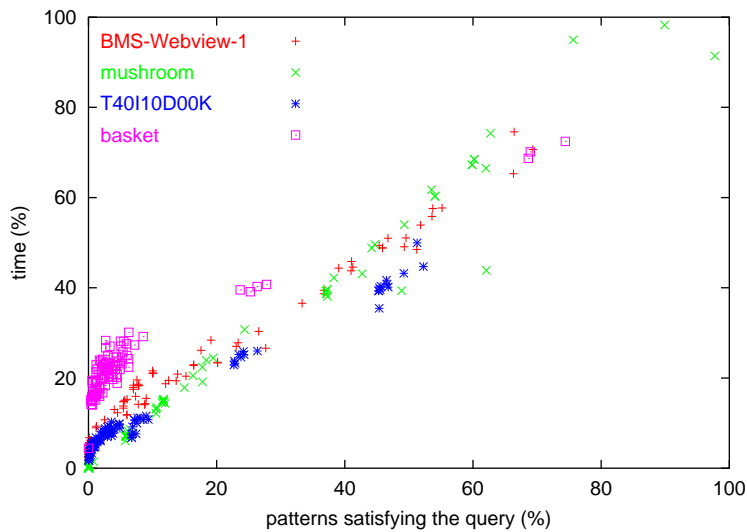}
\caption{Improvement after exploiting constraints.}
\label{improvement}
\end{figure}

For each data set, we generated $100$ random Boolean queries
consisting of at most three atomic conditions.
Figure~\ref{improvement} shows the improvement on the performance
of the algorithm exploiting the constraints. The y-axis shows the
time needed for the algorithm exploiting our queries, relative to
the time needed without exploiting the queries.  The x-axis shows
 the number of patterns satisfying the given query, relative to
the total number of patterns.
As can be seen, the time needed to generate all frequent sets and
association rules is proportional to the restrictiveness of the
constraints. Notice that the proportionality factor is 1.

\section{Interactive Mining}\label{interactive}

\subsection{Integrated Querying or Post-Processing?}

In the previous section, we have seen a way to integrate
constraints tightly into the mining of association rules. We call
this {\em integrated querying}. At the other end of the spectrum
we have {\em post-processing}, where we perform standard,
non-constrained mining, save the resulting itemsets and rules, and
then query those results for the constraints.

Integrated querying has the following two obvious advantages over
post-proc\-ess\-ing:

\begin{enumerate}
\item
Answering one single data mining query using integrated querying
is much more efficient than answering it using post-processing.
\item
It is well known that, by setting parameters such as minimal
support too low, or by the nature of the data, association rule
mining can be infeasible simply because of a combinatorial
explosion involved in the generation of rules or frequent
itemsets.  Under such circumstances, of course, post-processing is
infeasible as well; yet, integrated querying can still be
executed, if the query conditions can be effectively exploited to
reduce the number of itemsets and rules from the outset.
\end{enumerate}

However, as already mentioned in the Introduction, data mining
query language environments must support an interactive, iterative
mining process, where a user repeatedly issues new queries based
on what he found in the answers of his previous queries.  Now
consider a situation where minimal support requirements and data
set particulars are favorable enough so that post-processing is
not infeasible to begin with. Then the global, non-constrained
mining operation, on the result of which the querying will be
performed by post-processing, \emph{can be executed once and its
result materialized for the remainder of the data mining session}.

In that case, if the session consists of, say, 20 data mining
queries, these 20 queries amount to standard retrieval queries on
the materialized mining results.  In contrast, answering every
single of the 20 queries by integrated querying will involve at
least 20, and often many more, passes over the data, as each query
involves a separate mining operation. Also, several queries could
have a non-empty intersection, such that a lot of work is repeated
several times. Hence, the total time needed to answer the
integrated queries is guaranteed to grow beyond the
post-processing total time.

\begin{figure}
\centering
\includegraphics{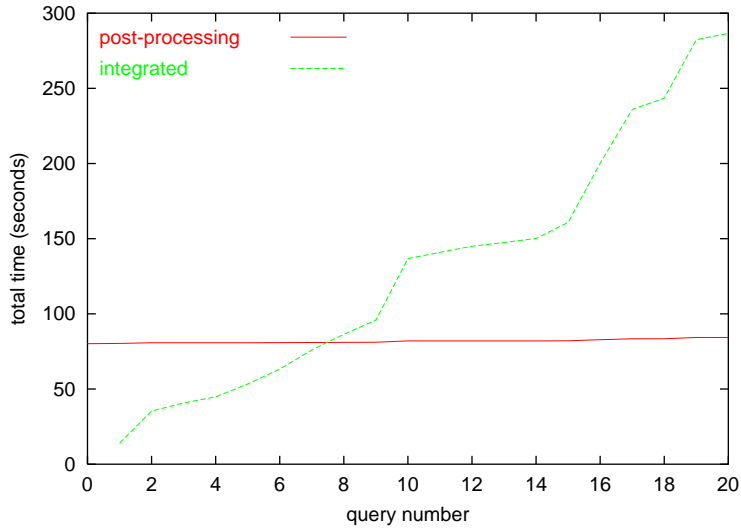}
\caption{Integrated querying versus post-processing.}
\label{cutoff}
\end{figure}

The naively conceived advantages of integrated querying over
post-processing become much less clear now.  Indeed, if the number
of data mining queries issued by the user is large enough, then
the post-processing approach clearly outperforms the integrated
querying approach. We have performed several experiments on the
data sets described in the previous section which all confirmed
this predicted effect.  However, for the post-processing approach,
we only materialized all frequent itemsets since the time needed
to generate all association rules that satisfy the query turned
out to be as fast as finding all such rules from the materialized
results. Figure~\ref{cutoff} shows the total time needed for
answering up to $20$ different queries on the BMS-Webview-1 data
set. Since the time needed to generate all association rules is
the same for both approaches, we only recorded the time to
generate all itemsets that were needed to generate all association
rules. The queries were randomly generated, only those queries
with an empty output were replaced, but all used the same support
threshold as was used for the initial mining operation of the
post-processing approach. As can be seen, the cut-off point from
where the post-processing approach outperforms the integrated
querying approach occurs already after the eighth query.

\subsection{Incremental Querying: Basic Approach}

From the above discussion it is clear that we should try to combine
the advantages of integrated querying and post-processing.
We now introduce such an approach, which we call \emph{incremental querying}.

In the incremental approach, all itemsets that result from every
posed query, as well as all intermediate generated itemsets, are
stored into a cache.  Initially, when the user issues his first
query, nothing has been mined yet, and thus we answer it using
integrated querying.

Every subsequent query is first converted to its corresponding
rule- and set-query in disjoint DNF. For every disjunct in the
set-query, the system adds all currently cached itemsets that
satisfy the disjunct to the data structure holding itemsets, that
is used for mining that disjunct, as well as all of its subsets
that satisfy the disjunct (note that these subsets may not all be
cached; if they are not, we have to count their supports during
the first scan through the data set). We also immediately add all
candidate itemsets.

If no new candidate itemsets can be generated, which means that
all necessary itemsets were already cached, we are done. However,
if this is not the case, we can now begin our queried mining
algorithm with the important generalization that in each
iteration, candidate itemsets of different cardinalities are now
generated. In order for this to work, candidate itemsets that turn
out to be infrequent must be kept such that they are not
regenerated in later iterations. This generalization was first
used by Toivonen in his sampling algorithm~\cite{sampling}.

Caching all generated itemsets gives us another advantage that can
be exploited by the integrated querying algorithm. Consider a
set-query stating that items $1$ and $2$ must be in the itemsets.
In the first iteration of the algorithm, all single itemsets are
generated as candidate sets over the new data set ${\cal D}_0$
(cf. Section~\ref{conjunct}). We explained that these single
itemsets actually represent supersets of $\{1,2\}$. Normally,
before we generate a candidate itemset, we check if all of its
subsets are frequent.  Of course, this is impossible if these
subsets do not even exist in ${\cal D}_0$. Now, however, we can
check in the cache for a subset with too low support; if we find
this, we avoid generating the candidate.

We thus obtain an algorithm which reuses previously generated
itemsets as if they had been generated in previous iterations of
the algorithm. We are optimal in the sense that we never generate
and test itemsets that were generated before. For rule generation,
we again did not cache the results, but in stead generated all
association rules when needed for the same reasons as explained in
the previous section.

In the worst case, the cached results do not contain anything that
can be reused for answering a query, and hence the time needed to
generate the itemsets and rules that satisfy the query is equal to
the time needed when answering that query using the integrated
querying approach. In the best case, all requested itemsets are
already cached, and hence the time needed to find all itemsets and
rules that satisfy the query is equal to the time needed for
answering that query using post-processing. In the average case,
part of the needed itemsets are cached and will then be used to
speed up the integrated querying approach.  If the time gained by
this speedup is more than the time needed to find the reusable
sets, then the incremental approach will always be faster than the
integrated querying approach. In the limit, all itemsets will be
materialized, and hence all subsequent queries will be answered
using post-processing.

\subsection{Incremental Querying: Overhead}

Could it be that the time gained by the speedup in the integrated
querying approach is less than the time needed to find and reuse
the reusable itemsets?  This could happen when a lot of itemsets
are already cached, but almost none of them satisfy the
constraints. It is also possible that the reusable itemsets give
only a marginal improvement. We can however counter this
phenomenon by estimating what is currently cached, as follows.

We keep track of a set-query $\phi_{sets}$ which describes the
stored sets. This query is initially {\it false}. Given a new
query (rule-query) $\psi$, the system now goes through the
following steps: (step~1 was described in Section~\ref{conjunct})
\begin{enumerate}
\item Convert the rule-query $\psi$ to the set-query $\phi$ \item
$\phi_{\it mine} := \phi \wedge \neg \phi_{\it sets}$ \item
$\phi_{\it sets} := \phi_{\it sets} \vee \phi$
\end{enumerate}
After this, we perform:
\begin{enumerate}
\item \label{prisets}
Generate all frequent sets according to
$\phi_{\it mine}$, using the basic incremental approach.
\item \label{postsets}
Retrieve all cached sets satisfying $\phi \wedge \neg \phi_{\it mine}$.
\item \label{subsets}
Add all needed subsets that can serve as bodies or heads.
\item \label{postrules}
Generate all rules satisfying $\psi$.
\end{enumerate}

Note that the query $\phi_{\it mine}$ is much more specific than
the original query $\phi$. We thus obtain a speedup, because we
have shown in Section~\ref{queries} that the speed of integrated
querying is proportional to the restrictiveness of the query.

\subsection{Avoiding Exploding Queries}

The improvement just described incurs a new problem.
The formula $\phi_{\it sets}$ becomes longer with the session.
When, given the next query $\phi$, we mine for $\phi \wedge \neg \phi_{\it sets}$,
and convert this to disjoint DNF which could explode.

To avoid this, consider $\phi_{\it sets}$ in DNF: $\phi_1 \vee \cdots \vee \phi_n$.
Instead of the full query $\phi \wedge \neg \phi_{\it sets}$,
we are going to use a query $\phi \wedge \neg \phi'_{\it sets}$,
where $\phi'_{\it sets}$ is obtained from $\phi_{\it sets}$ by keeping only
the least restrictive disjuncts $\phi_i$ (their negation will thus be most restrictive).
In this way $\phi \wedge \neg \phi'_{\it sets}$ is kept short.

But how do we measure restrictiveness of a $\phi_i$?
Several heuristics come to mind.
A simple one is to keep for each $\phi_i$ the number of cached sets that satisfy it.
These numbers can be maintained incrementally.

\subsection{Experiments}

\begin{figure}
\centering
\subfigure[basket]{\includegraphics{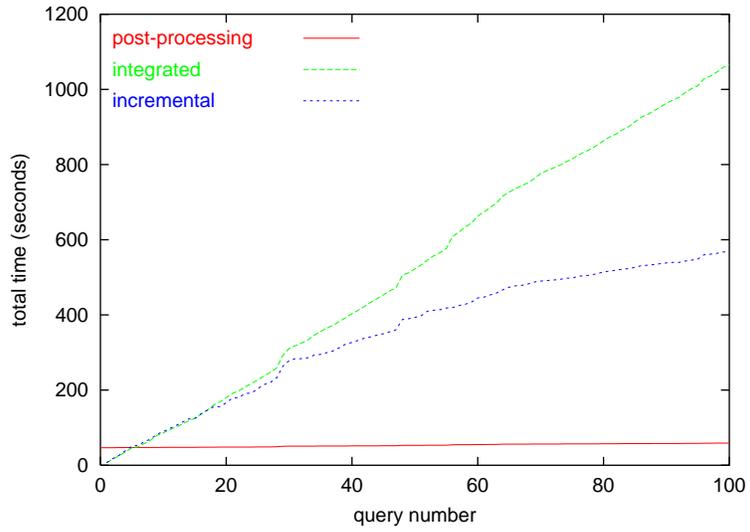}}\\
\subfigure[BMS-Webview-1]{\includegraphics{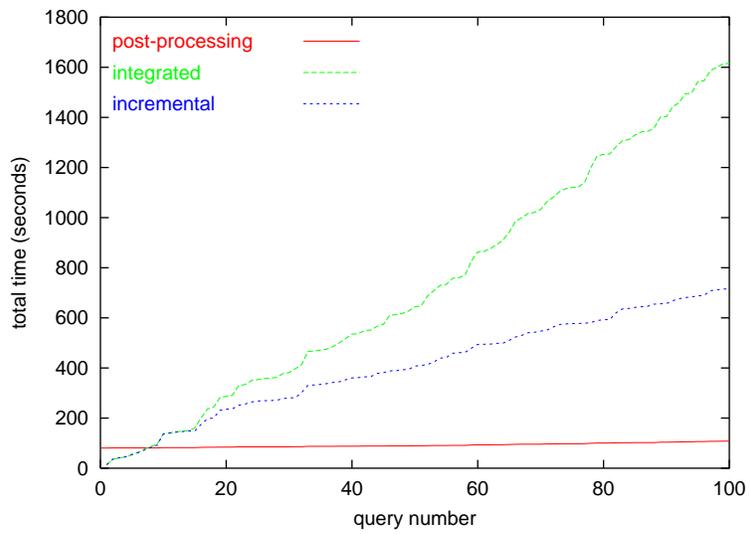}} \caption{Actual
and estimated number of candidate patterns.}
\label{fig:incremental}
\end{figure}

For each data set described in Section~\ref{queries} we
experimented with a session of $100$ queries using the integrated
querying approach, the post-processing approach and the
incremental approach. Again, the queries used for the sessions
where randomly generated. Figure~\ref{fig:incremental} shows the
evolution of the sessions in time.
\addtocounter{figure}{-1}
\begin{figure}
\addtocounter{subfigure}{2} \centering
\subfigure[T40I10D100K]{\includegraphics{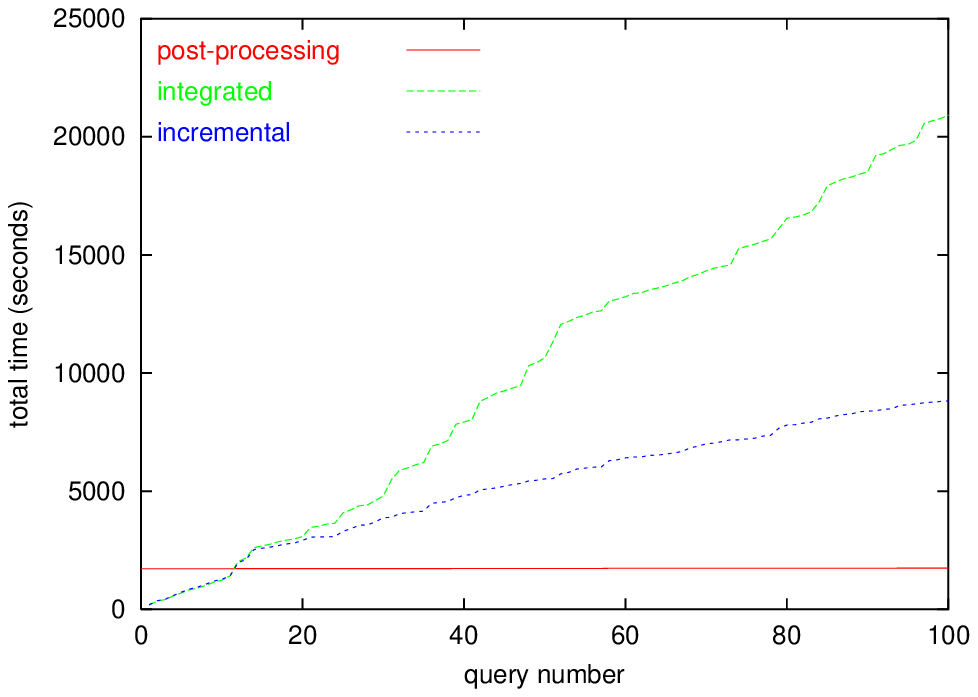}}\\
\subfigure[mushroom]{\includegraphics{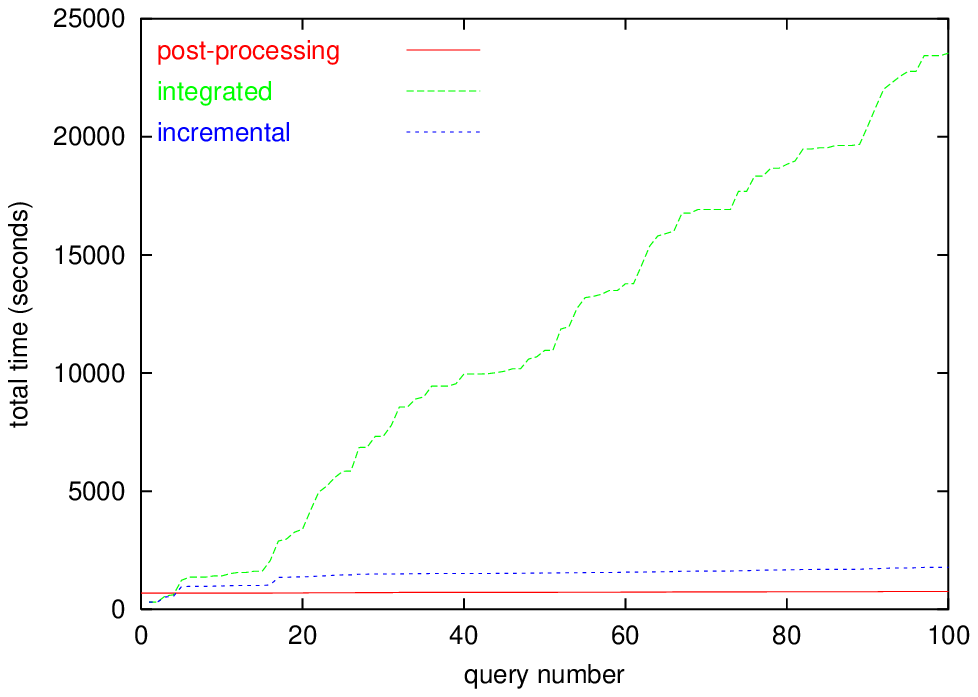}} \caption{Actual and
estimated number of candidate patterns.}
\end{figure}
For all four sessions, the cut-off point where the integrated
querying approach loses against the post-processing approach is
the same for the incremental querying approach since not enough
itemsets could be reused before that. Except for the mushroom data
set, the incremental approach starts paying off after the
twentieth query.  Nevertheless, the reuse of previous results does
not improve the performance enough for the incremental approach.
Indeed, the incremental approach will always need some time to
fetch all pre-generated itemsets and it will try to generate some
more.  However, as can be seen, the incremental approach shows a
significant improvement on the integrated querying approach. Only
for the mushroom data set, the cut-off point occurs at the fifth
query, and almost all itemsets have been generated after the
eighteenth query. As can be seen, the performance of the
post-processing approach is very good compared to the other
approaches. Nevertheless, if we still lowered the support
thresholds, the post-processing approach becomes unfeasible to
begin with, due to an overload of frequent itemsets.  In that
case, the integrated and incremental approach are still feasible
and perform very similar as in the presented experiments.

\section{Conclusions} \label{conclude}

This study revealed several insights into the association rule
mining problem. First, due to recent advances on association rule
mining algorithms, the performance has been significantly
improved, such that the advantages of integrating constraints into
the mining algorithm suddenly become less clear.  Indeed, we
showed that as long mining without any constraints is feasible,
that is, if the number of frequent itemsets does not reach a huge
amount, the total time spent to query the frequent itemsets and
confident association rules becomes less after a certain amount of
queries, compared to integrated querying, in which every query is
pushed into the mining algorithm.  The incremental approach still
improves the integrated approach by reusing as much previously
generated results as possible. If the cut-off point would lie
beyond the number of queries in which the user is interested, the
incremental approach is obviously the best choice to use.

Of course, if the user is interested is some frequent itemsets and
association rules which have very low frequencies, and hence
mining without any constraints becomes infeasible, the incremental
approach can still be performed.

Also note, that if a user is still interested in all frequent sets
and association rules, but mining without constraints is
infeasible, our queries can be used to divide the task over
several runs, without spending much more time.  For example, one
can ask different queries of which the disjunction still gives all
sets and rules. Essentially, this technique forms the basis of the
well known Eclat~\cite{eclat} and FP-growth
algorithms~\cite{fpgrowth}.

\section*{Acknowledgement}

We wish to thank Blue Martini Software for contributing the KDD
Cup 2000 data, the machine learning repository librarians 
Catherine Blake and Chris Mertz for providing access to the 
mushroom data, and Tom Brijs 
for providing the Belgian retail market basket data.

\bibliographystyle{plain}

\end{document}